\documentclass[prl,twocolumn,superscriptaddress,floatfix,noshowpacs,notitlepage]{revtex4-1}%
\usepackage{graphicx,bm,times}
\usepackage{amsmath}
\usepackage{amsfonts}
\usepackage{amssymb}
\usepackage{color}

\begin{document}

\title{Multiband one-dimensional electronic structure and spectroscopic signature of Tomonoga-Luttinger liquid behavior in K$_2$Cr$_3$As$_3$}

\author{M. D. Watson}
\email[corresponding author:]{matthew.watson@diamond.ac.uk}
\affiliation{Diamond Light Source, Harwell Campus, Didcot, OX11 0DE, UK}
\author{Y. Feng}
\affiliation{State Key Laboratory of Surface Physics and Department of Physics, Fudan University, Shanghai 200433, China}
\author{C. W. Nicholson}
\affiliation{Department of Physical Chemistry, Fritz-Haber-Institut of the Max Planck Society, Faradayweg 4-6, Berlin 14915, Germany}
\author{C. Monney}
\affiliation{Department of Physics, University of Zurich, Winterthurerstrasse 190, 8057 Zurich, Switzerland}
\author{J. M. Riley}
\affiliation{Diamond Light Source, Harwell Campus, Didcot, OX11 0DE, UK}
\affiliation{SUPA, School of Physics and Astronomy, University of St Andrews, St Andrews, Fife KY16 9SS, UK}
\author{H.~Iwasawa}
\affiliation{Diamond Light Source, Harwell Campus, Didcot, OX11 0DE, UK}
\author{K.~Refson}
\affiliation{Department of Physics, Royal Holloway, University of
	London, Egham, Surrey TW20 0EX, UK}
\author{V.~Sacksteder}
\affiliation{Department of Physics, Royal Holloway, University of
	London, Egham, Surrey TW20 0EX, UK}
\author{D.~T.~Adroja}
\affiliation{ISIS Facility, Rutherford Appleton Laboratory, Chilton, Didcot, Oxon, OX11 0QX, UK}
\author{J. Zhao}
\affiliation{State Key Laboratory of Surface Physics and Department of Physics, Fudan University, Shanghai 200433, China}
\author{M. Hoesch}
\affiliation{Diamond Light Source, Harwell Campus, Didcot, OX11 0DE, UK}

\begin{abstract}
We present Angle-Resolved Photoemission Spectroscopy measurements of the quasi-one dimensional superconductor K$_2$Cr$_3$As$_3$. We find that the Fermi surface contains two Fermi surface sheets, with linearly dispersing bands not displaying any significant band renormalizations. The one-dimensional band dispersions display a suppression of spectral intensity approaching the Fermi level according to a linear power law, over an energy range of $\sim$ 200~meV. This is interpreted as a signature of Tomonoga-Luttinger liquid physics, which provides a new perspective on the possibly unconventional superconductivity in this family of compounds. 
\end{abstract}
\date{\today}
\maketitle


The recently-discovered quasi-one dimensional (q1D) superconductors A$_2$Cr$_3$As$_3$ (A=K,Rb,Cs) present a new opportunity for the study of the phenomena which may occur when electrons are effectively confined to one dimension.  The 
hexagonal crystal structure consists of infinite  stacks of  [Cr$_3$As$_3$]  clusters running along the $c$ axis, which form the one-dimensional structural motif~\cite{Bao2015,Cao2016}. The A$^+$ ions between the chains transfer charge and mediate inter-chain coupling thus rendering the material q1D. The inclusion of Cr in the structure is a rare example of incorporating $3d$ orbitals into a q1D system, as opposed to the well-known case of $4d$ orbitals in metallic Mo-based q1D systems Li$_{0.9}$Mo$_6$O$_{17}$ and Tl$_2$Mo$_6$Se$_6$, which brings the possibility of strong correlations \cite{Bao2015} and magnetic interactions, which may be frustrated due to the triangular Cr motifs in the structure \cite{Wu2015}. Superconductivity is found at 6.1 K in K$_2$Cr$_3$As$_3$ \cite{Kong2015}, which ranks amongst the highest $T_c$ for q1D systems, and has been suggested to be of an unconventional nature \cite{Bao2015} with some experimental \cite{Adroja2015,Liu2016} and theoretical \cite{Zhong2015,Wu2015} support for a triplet pairing state. Finally the one-dimensionality of the system makes it a candidate for Tomonaga-Luttinger liquid (TLL) physics \cite{Zhi2015,Zhong2015,Miao2016_arxiv}. 
This rich set of phenomena constitutes a very strong motivation for experimental determinations of the electronic structure by Angle-Resolved Photoemission Spectroscopy (ARPES) measurements.  

In this Letter we present a detailed study of the electronic structure of single crystals of K$_2$Cr$_3$As$_3$ using  ARPES. We find a Fermi surface containing two linearly dispersing hole-like q1D Fermi surface sheets. The overall bandwidth of the Cr $3d$ bands and Fermi velocities are comparable to DFT calculations, indicating that the correlated Fermi liquid picture is not appropriate for K$_2$Cr$_3$As$_3$. Furthermore we show that the spectral weight of the q1D bands decreases according to an approximately linear power law over a $\sim$200 meV energy scale up to the Fermi level, also obeying a universal temperature scaling relation. This power-law depletion of spectral weight is a signature of TLL behavior in photoemission measurements of one-dimensional systems. We conclude that the wide range of fascinating experimental results on A$_2$Cr$_3$As$_3$ should be interpreted within the framework of a q1D system close to TLL physics. 

\begin{figure*}
	\centering
	\includegraphics[width=0.95\linewidth]{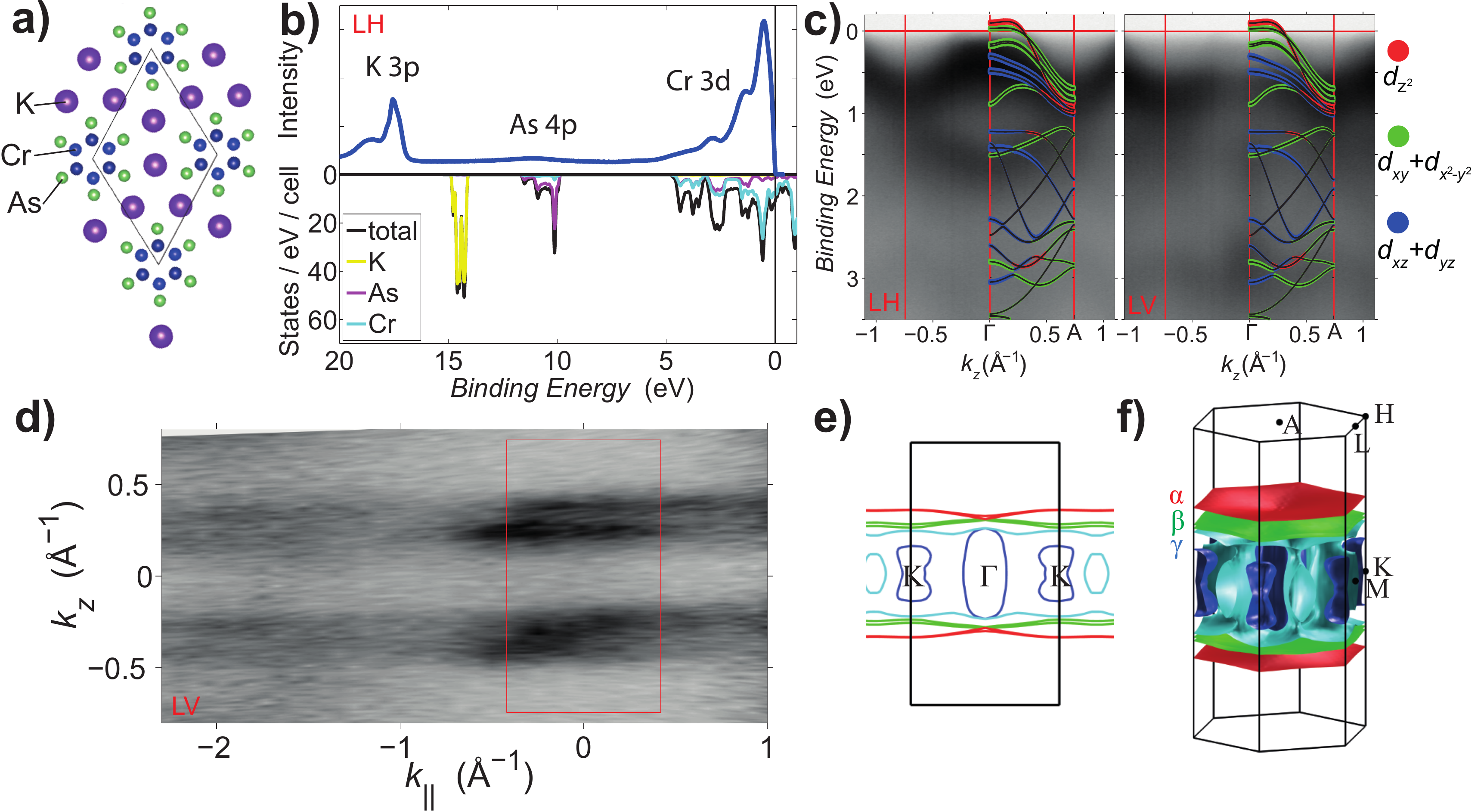}
	\caption[Fig1]{a) Crystal structure of K$_2$Cr$_3$As$_3$ viewed down the $c$-axis. b) Angle-integrated photoemission spectrum obtained at 10~K, compared with the Density of States calculated from DFT. The experimental and calculated Cr 3d bandwidth are in good agreement. c) ARPES measurements of Cr $3d$ valence band dispersions, obtained with horizontal (LH) and vertical (LV) linear polarizations of the incident photons. The spectra are overlaid with DFT calculations along the $\rm \Gamma$-A for reference, which show broad agreement. The calculated bands are colored by the primary orbital character of the band as indicated. d) Experimental Fermi surface map of K$_2$Cr$_3$As$_3$ constructed by integrating spectral weight within 50~meV of the Fermi level, consisting of two pairs of q1D bands. e,f) Calculated Fermi surface of K$_2$Cr$_3$As$_3$; here spin-orbit coupling is included and colours represent different bands not orbital characters.}
	\label{fig1}
\end{figure*}

\begin{figure}
	\centering
	\includegraphics[width=0.95\linewidth]{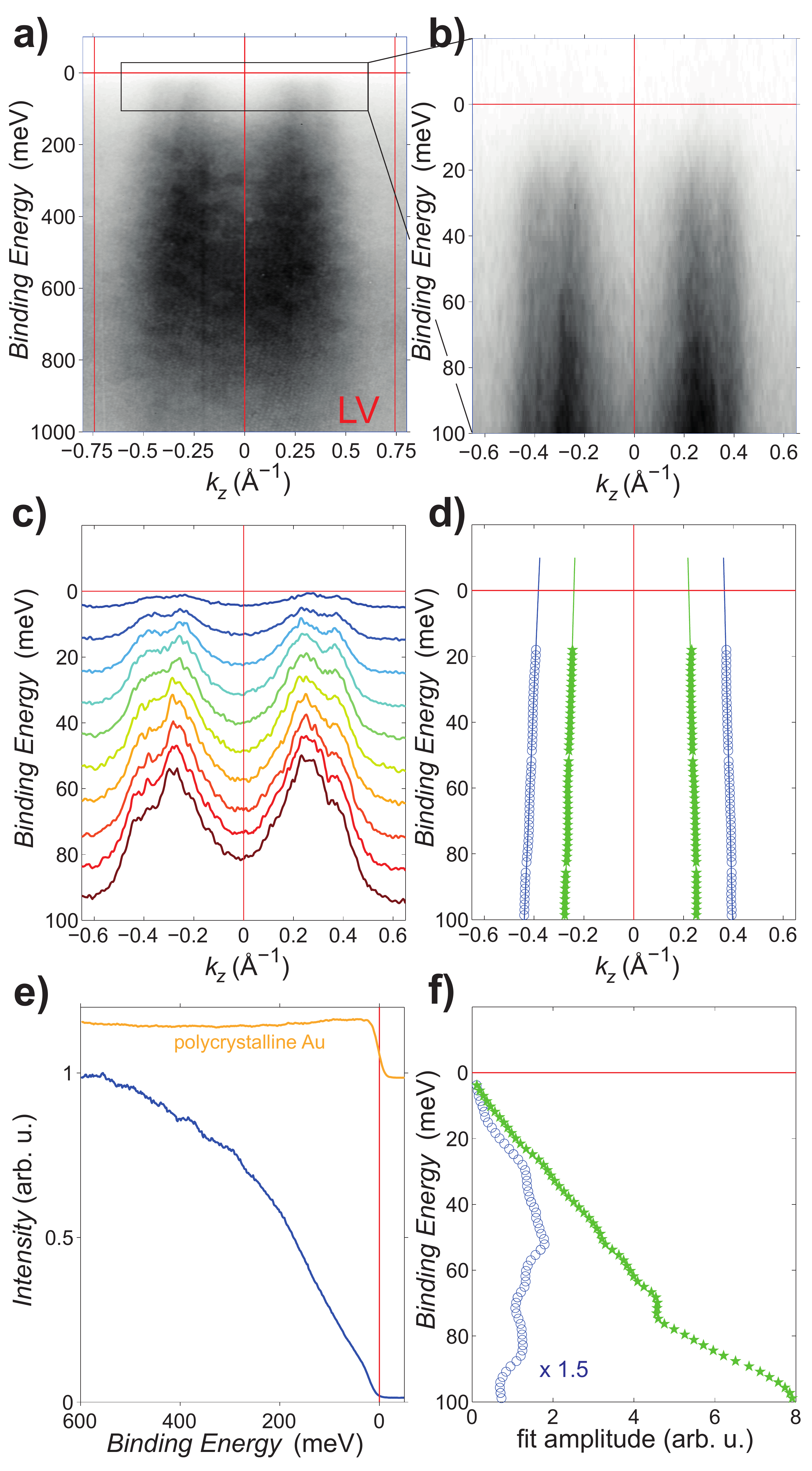}
	\caption{a) ARPES dispersion obtained at 10~K for a cut along the $k_z$ direction. Vertical red lines indicate the Brillouin zone centre and boundary. b) Detailed dispersions close to the Fermi level and c) selected slices of raw data, showing the suppression of spectral weight approaching the Fermi level. d) Band positions and f) fit amplitude of the outer and inner bands. e) Integrated spectral weight for this cut, compared to a polycrystalline Au reference.}
	\label{fig2}
\end{figure}


The highly pure K$_2$Cr$_3$As$_3$ single crystals were prepared using a high temperature solution growth method \cite{Bao2015,Kong2015}. The starting materials were loaded in an alumina crucible following the ratio of K: Cr: As = 6:1:7. The alumina crucible was welded in a Ta tube and then sealed in an evacuated quartz ampoule. The quartz ampoule was slowly heated up to 1173~K and held for 24 h in a furnace, followed by cooling down to 923 K at a rate of 2 K/h. ARPES measurements were performed at the I05-ARPES beamline at Diamond Light Source, UK. Straight stick-like samples with typical sizes of $\sim$2000x200x200$\mu$m were selected and prepared for ARPES measurements with the standard top-post cleavage method in ultrahigh vacuum (UHV). The gluing of the samples to the holders and the attachment of top posts was performed in a dry Ar glove box. Samples were carefully aligned such that the long axis of the stick-like samples was parallel with the detector, enabling cuts in $k_z$ (dispersion parallel to the crystal $c$-axis) to be obtained. In this geometry, varying either polar rotation of the sample or the incident photon energy manipulates the wavevector component $k_{||}$ in the $(a,b)$-plane of the crystal, although absolute determination of the in-plane wavevector is difficult. Data are obtained with an incident photon energy of $h\nu=96$~eV, with linear polarization either along the crystal $c$-axis (``LV") or orthogonal across (``LH"). The experimental energy resolution was set to 23 meV. The Fermi level was obtained from reference measurements on polycrystalline gold. Density Functional Theory (DFT) calculations were performed in Wien2k~\cite{wien2k} using the experimental crystal structure \cite{Bao2015}. 

In Fig.~\ref{fig1}a), we show the crystal structure of K$_2$Cr$_3$As$_3$ projected in the (a,b) plane \cite{Bao2015}. The Cr$_3$As$_3$ chains extend along the $c$ axis are separated by K$^{+}$ guest ions in two inequivalent positions. Thus the electronic structure is expected to be dominated by q1D dispersions arising mainly from Cr $3d$ orbitals, which combine into molecular orbitals on the triangular Cr plaquettes~\cite{Alemany2015,Zhong2015} . In Fig.~\ref{fig1}b) we present the shallow core level spectrum of K$_2$Cr$_3$As$_3$. The K $3p$ core levels around 18~eV display a multi-peak structure \footnote{The disagreement between the experimental location of the K $3p$ core level and the DFT calculation is an artifact related to the the often problematic calculation of shallow core levels in DFT}, reflecting the  inequivalent K sites~\cite{Bao2015} . A shallow bump around 11 eV binding energy matches well with the As $4p$ levels in the calculation. The brightest feature in the spectrum from 4~eV up to the Fermi level consists of the Cr $3d$ valence bands. Fig.~\ref{fig1}c) shows the dispersions of these bands along $k_z$. The observed periodicity along $k_z$ matches with the expected Brillouin zone, and some dispersions can be associated with particular bands in the DFT calculations. Thus we are confident that our measurements are representative of the bulk band dispersions derived from various subshells of the Cr~$3d$-band and that the sample is a single crystal. Due to photoemission matrix element effects, the intensity of dispersions depends strongly on the light polarization. In particular, the intense and highly dispersive band in the left panel of Fig.~\ref{fig1}c) (LH polarization) is suppressed in LV polarization (right panel), whereas the q1D bands extending up to the Fermi level are primarily observed in LV polarization. We note that these spectra were taken from a freshly cleaved sample; over a relatively short timescale of a few hours in good UHV the Cr $3d$ bands lose intensity and the detail in the K core level structure is lost as the surface degrades (see Supplemental Material, SM \footnote{See Supplemental Material at [URL will be inserted by publisher] for further scaling plots and analysis, details of the normalisation procedure, and a sample comparison.}).   

There is a good agreement between the DFT calculation and the experimental bandwidth of the Cr $3d$ bands, as shown in Fig.~\ref{fig1}b,c). This is a somewhat surprising observation given that K$_2$Cr$_3$As$_3$ has previously been suggested as a strongly correlated electron system due to the anomalously large Sommerfeld coefficient \cite{Bao2015,Kong2015}. Metallic systems with strong correlations usually manifest in ARPES with renormalized quasiparticle bands which become sharp at the Fermi level, and the overall bandwidth can appear compressed compared to DFT calculations. However here the observed Cr $3d$ dispersions generally have a similar bandwidth to the calculations, and Fermi velocities are also similar as shown in Table~\ref{tab1}, thus only minor or no renormalization effects are found.

In Fig.~\ref{fig1}d) we present the Fermi surface map of K$_2$Cr$_3$As$_3$. This map is constructed by integrating the spectral weight to a much deeper binding energy of 50~meV than would routinely be used in ARPES analysis; this is necessary since the measured spectral weight disappears as the Fermi level is approached, as will be discussed later. The DFT calculations shown in Fig.~\ref{fig1}e,f) predict the existence of two pairs of q1D bands ($\alpha$,$\beta$) and a q3D band ($\gamma$); due to the absence of inversion symmetry in the crystal structure \cite{Bao2015} the inclusion of spin-orbit coupling in the calculation splits these bands further. Experimentally, the Fermi surface is found to include only two 1D bands, with no indication of any 3D Fermi surfaces. Any spin-orbit induced splitting is not resolved.

\begin{table}[b]
	\centering
	\caption{Dispersion parameters of the q1D bands $\alpha$ and $\beta$. $v_F$ is estimated from fits to DFT dispersion over the same range as the experimental data for direct comparison.}
	\label{tab1}
	\begin{tabular}{l|l|lll}
		& Exp ~~~~~   & $\Gamma$-A~~~ & K-H~~~ & M-L~~~ \\ \hline
		$k^\alpha_F$ (\AA$^{-1}$) & 0.37(4) &  0.3           & 0.35    &  0.35   \\ 
		$v^\alpha_F$ (eV \AA)    & 2.4(8) &    1.3            & 2.6    &  2.8   \\
		$k^\beta_F$ (\AA$^{-1}$)  & 0.23(4) &  0.27              & 0.28    & 0.28    \\
		$v^\beta_F$ (eV \AA)     & 2.8(8) &    1.7            &  1.3   &   1.3 
	\end{tabular}
\end{table}

In Fig.~\ref{fig2} we present a detailed analysis of cut a measured in LV polarization which includes the pair of bands which approach the Fermi level. Band positions are extracted in Fig.~\ref{fig2}d) by constrained fits of pairs of Lorentzian lineshapes. Both band dispersions are found to be near linear over this energy range, and the derived Fermi wavevectors and velocities are summarized in Table \ref{tab1}. The most notable feature of this spectrum is that the spectral weight decreases gradually as the Fermi level is approached, which is clear from both from the data slices in Fig.~\ref{fig2}c) and the integrated spectral weight in Fig.~\ref{fig2}e). In Fig.~\ref{fig2}f) we plot the intensity of the bands as derived from the peak fitting. Both bands show a near linear rise of intensity within $E_B < 0.03$~eV, after which the outer band intensity saturates to finite values, while the inner band, which dominates in the total intensity, continues to rise.

While a resolution-limited Fermi cut-off is typically observed in ARPES measurements, as shown by the polycrystalline Au reference measurement in Fig.~\ref{fig2}e), the continuous suppression of the single-particle spectral weight is a characteristic of ARPES measurements of several different q1D systems~\cite{dardel91}. One could consider if it may result from order formation, such as a charge density wave leading to a gap opening at $E_F$. However, given that our momentum space surveys show the same spectral weight depletion everywhere in momemtum space, while the material exhibits metallic resistivity without any anomalies except the superconducting transition \cite{Kong2015,Liu2016}, we exclude an ordering-related energy gap at $E_F$. The interplay between disorder and electron interactions may also deplete the DOS near the Fermi level \cite{Richardella2010,altshuler1979}, which has recently been proposed for the interpretation of ARPES spectra \cite{yaji2016_arxiv,kobayashi2007}. In metallic systems, the Altshuler-Aronov theory \cite{altshuler1979} predicts that density of states near the Fermi level is a non-zero constant plus an anomalous power law contribution (in 3-D) \cite{kobayashi2007}. However we can dismiss this assignment since our measured DOS extrapolates to zero at the Fermi level even though the samples are good conductors. 

Thus we interpret the missing spectral weight on these q1D dispersions as evidence for the peculiar excitation spectra of a Tomonoga-Luttinger liquid (TLL), where in one dimension the quasi-particle description of excitations breaks down and the highly collective modes allow for low energy excitations only at infinitely small momentum transfer. This leads to a vanishing quasiparticle weight $Z$~\cite{Giarmarchi2004}, and a recovery of spectral weight according to a power law, both as a function of binding energy and temperature. Spectra in agreement with the power law predictions of this model have been observed in varied q1D systems including Li$_{0.9}$Mo$_6$O$_{17}$ \cite{Wang2006pb}, carbon nanotubes~\cite{ishii03} and Bi chains grown on InSb(001)~\cite{Ohtsubo2015}.

\begin{figure}
	\centering
	\includegraphics[width=\linewidth]{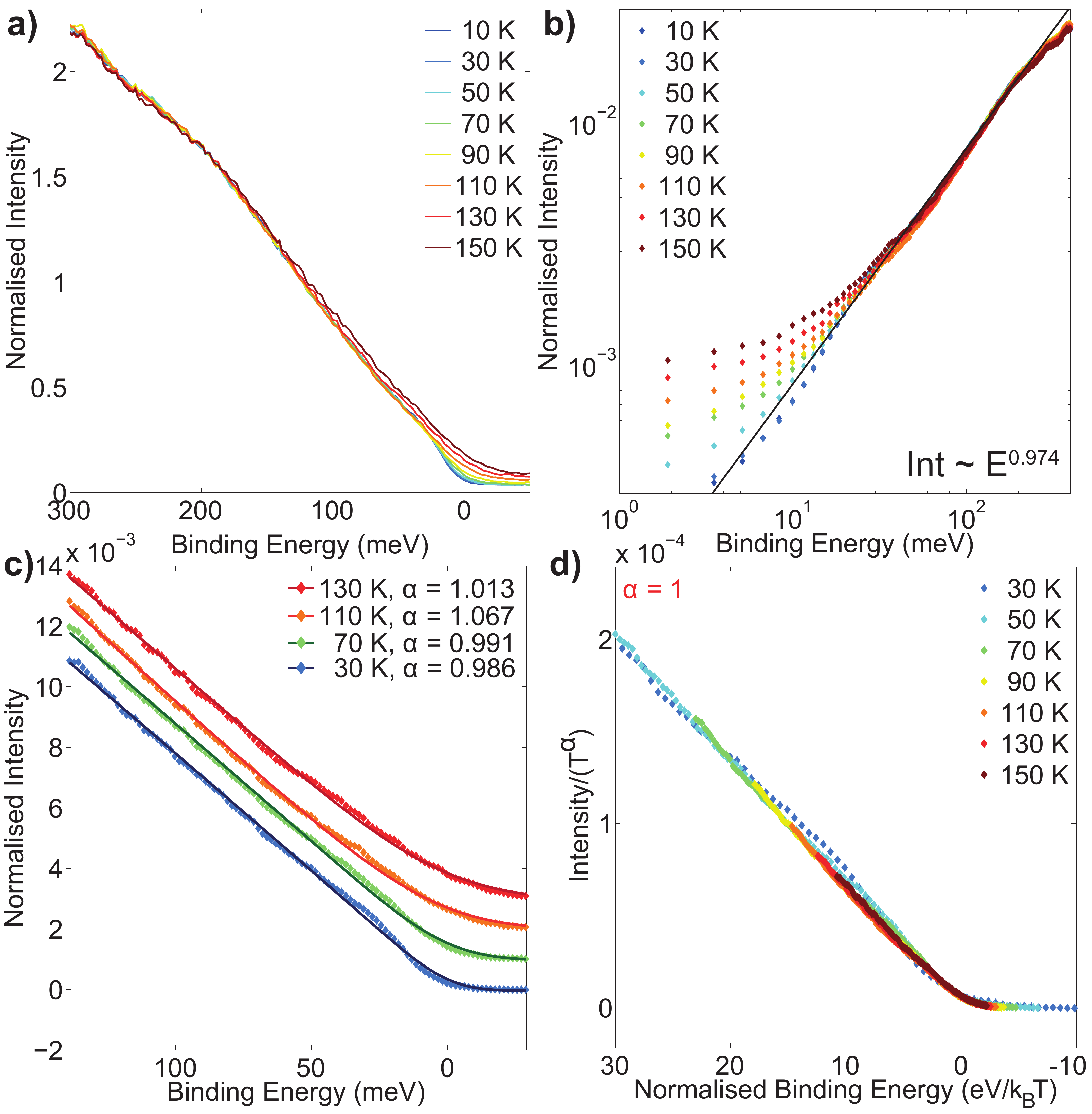}
	\caption[tdep]{Integrated spectral weight within in the range $-0.5 < k_z < 0.5$~\AA$^{-1}$. Data are normalised in the range $250 <E_B < 200$~meV. b) Logarithmic plot of (a), after subtraction of constant offset from background (taken well above $E_F$). Black line is linear fit of logarithmic data in range $20 <E_B < 200$~meV at 10~K, implying a power law of $\alpha=0.974$. c) Fits to TLL spectral function at selected temperatures, obtaining very similar values of $\alpha$. Data are offset for clarity. d) Scaling plot, showing that the data at various temperatures collapse onto one curve for $\alpha=1$.}
	\label{fig3}
\end{figure}

In Fig.~\ref{fig3}a) we present the temperature-dependence of integrated intensity corresponding to the data shown in Fig.~\ref{fig2}. The power law behavior of the spectral weight at low temperatures is directly visualized in the double-logarithmic plot of Fig.~\ref{fig3}b). A fit to the data at 10~K yields a nearly linear ($\alpha$=0.974) exponent which applies over an energy range of more than an decade. At higher temperatures, the data depart from the simple power law curve below $\sim$30~meV, and a more substantial spectral weight remains at the Fermi level. To model the finite temperature effects, we use a more detailed formula for the intensity ($I$) of the spectral function, which has a smooth transition from a low-energy regime involving the Fermi-Dirac distribution function $f(\epsilon,T)$ to a high-energy regime dominated by a power law relation: 
\begin{equation}
I(\epsilon,T)\propto T^\alpha \cosh \left( \frac{\epsilon}{2}\right ) \left | \Gamma \left ( \frac{1+\alpha}{2} + i\frac{\epsilon}{2\pi} \right ) \right |^2 f(\epsilon,T), 
\label{eq1}
\end{equation}
where $\epsilon=E/{k_BT}$ (cited from~\cite{Ohtsubo2015}, see also Refs.~\cite{Schonhammer1993,Markhof2016} for further discussion).

Direct fits of this equation (convolved with experimental resolution) to data at selected temperatures in the binding energy range $140<E_B< -30$~meV are presented in Fig.~\ref{fig3}c), showing good agreement. The fitted exponents $\alpha$ are found to be in the range 0.97-1.03 for all measured temperatures. Eq.~(\ref{eq1}) also implies a scaling relation; on a plot of $I/{T^\alpha}$ as a function of temperature-renormalized energy $\epsilon=E/{k_BT}$, the various curves as a function of temperature should collapse for the correct choice of $\alpha$. The plot with $\alpha=1$ in Fig.~\ref{fig3}d) shows excellent scaling behavior; other choices of $\alpha$ away from 1 show progressively worse scaling (SM). The combination of the power law fit, the fitting of Eq.~\ref{eq1} and the temperature scaling relation gives good evidence that K$_2$Cr$_3$As$_3$ follows the expected behavior for a TLL on both the temperature and energy axes, with a characteristic exponent $\alpha\simeq$1. The temperature-independent exponent $\alpha\simeq$1 is of a comparable magnitude to exponents observed in other recognized TLL systems~\cite{allen1995,Gweon2003,Wang2006}.  A decisive proof of the TLL state could come from the observation of spin-charge separation. Our data do not provide such evidence, though it may be hidden in the considerably large widths of the features as sharp spinon and chargon excitations are only expected in cases with much lower $\alpha <0.5$~\cite{Gweon2003}. We thus conclude that the photoelectron spectroscopy data presented here constitute evidence for TLL or closely related physics.

Resistivity measurements along the $c$ axis of clean single crystal samples are found to obey a $T^3$ scattering law \cite{Kong2015,Liu2016,Cao2016}, as opposed to the $T^2$ expectation for a Fermi liquid. Moreover NMR measurements of K$_2$Cr$_3$As$_3$ found an unusual temperature-dependence of the spin relaxation rate, inconsistent with the Korringa law expectation of a Fermi liquid but reminiscent of other TLL systems \cite{Zhi2015}. While spin fluctuations could still be relevant to both these discussions, deviation from Fermi liquid behavior is certainly expected in the context of these ARPES results which show an absence of low-energy quasiparticle excitations.  The ARPES results also demand a different explanation of the anomalously large Sommerfeld coefficient \cite{Kong2015,Bao2015}, other than the picture of quasiparticles renormalized by electron-electron (or electron-phonon \cite{Subedi2015,Zhang2015c}) interactions. We predict that STM studies would show a similar power law behavior to our observations \cite{Hager2005}. 

Our ARPES data shows that the TLL persists down to 10~K, near  the superconducting $T_c$.  Therefore the best starting point for understanding superconductivity may be  with  TLLs hosted by 1-D Cr$_3$As$_3$ stacks \cite{Zhong2015,Miao2016_arxiv} and only weakly coupled by the K guest ions, rather than with quasiparticles occupying 3-D Fermi surfaces \cite{Zhou2015,Wu2015}.  In q1D superconductors  the guest ion type and disorder are crucial for controlling superconductivity, as exemplified in the M$_2$Mo$_6$Se$_6$ family with guest ion M$=$Tl, In, Na, K, Rb  \cite{ansermet2016, petrovic2016,petrovic2010}.  Further ARPES and STM studies of the A$_x$Cr$_3$As$_3$ family, changing the guest ions  to Rb$_2$, Cs$_2$, K \cite{Tang2015,Tang2015b,Bao2015a}, could illuminate its superconducting state.

In conclusion, we have presented an ARPES study of single crystals of K$_2$Cr$_3$As$_3$. We find a Fermi surface containing two hole-like q1D Fermi surface sheets with linear dispersions. The spectral weight of the q1D bands decreases according to an approximately linear power law over a $\sim$200 meV energy scale up to the Fermi level, and obeys a universal temperature scaling. This is interpreted as a signature of TLL behavior. These results give a new perspective on the rich physics in the q1D A$_2$Cr$_3$As$_3$ superconductor family.

\begin{acknowledgments}
	\section{acknowledgments}
	We thank T.~Forrest, D.~Voneshen, A.P.~Petrovi\'{c}, Y.~Ohtsubo, and C.~Berthod for useful discussions. We thank T.~K.~Kim and D.~Daisenberger for technical assistance. We thank Diamond Light Source for access to Beamline I05 (Proposals No. SI13797 and NT14572) that contributed to the results presented here. C.M. acknowledges support by the Swiss National Science Foundation under grant number PZ00P2\_154867. J.Z. and Y.F. were supported by the National Natural Science Foundation of China (Grant No. 91421106) and the National Key R\&D Program of the MOST of China (Grant No. 2016YFA0300203).  J.M.R. acknowledges support from the EPSRC (grant number EP/L505079/1). 
\end{acknowledgments}



%

\end{document}